\documentclass[a4paper]{article}
\usepackage{amsmath}
\usepackage{amsfonts}
\usepackage{enumerate}
\usepackage{cite}

\normalbaselines
\begin{document}
\title{ SUSY partners of the truncated oscillator,
        Painlev\'e transcendents and B\"acklund transformations }
\author{DJ Fern\'andez C$^1$ and VS Morales-Salgado$^2$\\
Departamento de F\'{\i}sica, Cinvestav, A.P. 14-740, \sl 07000 M\'exico D.F.  Mexico\\
{\small$^1$david@fis.cinvestav.mx, $^2$vmorales@fis.cinvestav.mx}}

\maketitle

\begin{abstract}
 In this work the supersymmetric technique is applied to the truncated oscillator 
 to generate Hamiltonians ruled by second and third-order polynomial Heisenberg algebras, 
 which are connected to the Painlev\'e IV and Painlev\'e V equations respectively. 
 The aforementioned connection is exploited to produce particular solutions to both 
 non-linear differential equations and the B\"acklund transformations relating them.
\end{abstract}

\section{Introduction}
 Supersymmetric quantum mechanics (SUSY QM) is a technique typically used to modify the energy spectrum 
 of a given initial Hamiltonian $H_0$ \cite{wi81,wi82,mi84,ais93,aicd95,cks95,bs97,fgn98,fhm98,jr98,mnn98,
 qv99,sa99,mnr00,crf01,ast01,mr04,cfnn04,ff05,cf08,ma09,fe10,qe11,bf11a,ma12,ai12,ggm13,fm14,fm15}, 
 yielding new SUSY partners whose spectra are similar to the initial one. 
 Moreover, if the system of departure is ruled by certain algebraic structure, 
 then its SUSY partners turn out to be described by polynomial deformations of such an algebra. 
 
 In a recent paper, the SUSY transformations of first and second-order were applied to the harmonic oscillator 
 with an infinite potential barrier at the origin \cite{fm14}, which we will call truncated oscillator. 
 The underlying algebraic structure for the SUSY partners of the truncated oscillator was as well analyzed, 
 becoming a deformation of the Heisenberg-Weyl algebra known as polynomial Heisenberg algebra \cite{cfnn04,mn08,fh99,fnn04}.
 
 It is important to note that general systems ruled by second- and third-order polynomial Heisenberg algebras, 
 described by one-dimensional Schr\"odinger Hamiltonians having third and fourth-order differential ladder operators, 
 are linked to the Painlev\'e IV and V equations respectively 
 \cite{mn08,sh92,vs93,ad94,dek94,ekk94,ain95,srk97,acin00,bfn14,bf11b,be12}. 
 This connection was exploited in \cite{fm14} to generate solutions to the Painlev\'e IV equation, 
 using the SUSY partner Hamiltonians for the truncated oscillator, 
 where it was used that the initial system is ruled by the Heisenberg-Weyl algebra. 
 
 Since the truncated oscillator is also characterized by a first-order polynomial Heisenberg algebra, 
 which second-order generators act on the energy eigenstates more naturally than those of the Heisenberg-Weyl algebra, 
 thus there are SUSY partner Hamiltonians having as well fourth-order differential ladder operators and, 
 consequently, being linked with the Painlev\'e V equation. 
 Here we continue the work done in \cite{fm14} in the aforementioned natural direction.
 First we revisit the SUSY partners of the truncated oscillator, 
 considering now that they are ruled by both, second and third-order polynomial Heisenberg algebras. 
 Moreover, the corresponding solutions to the Painlev\'e IV and V equations will be studied, 
 along with the B\"acklund transformations which will clarify the way in which these solutions are related among each other. 

 With such a goal in mind, the article has been organized as follows. 
 In Section 2 the supersymmetric partners of the truncated oscillator are 
 obtained using first- and second-order transformations. 
 The connection between these SUSY partners and PIV and PV equations is described in Sections 3 and 4, 
 respectively, along with a straightforward procedure to obtain solutions to these non-linear equations. 
 The Backl\"und transformations relating these Painlev\'e transcendents are found in Section 5. 
 The conclusions and final remarks of the research can be found in Section 6.
 
\section{Supersymmetric partners of the truncated oscillator}
 As was stated in the introduction, the system on which the whole procedure 
 will be developed is the truncated oscillator, described by the 
 Hamiltonian $H_0=-\frac{1}{2}\frac{{\rm d}^2}{{\rm d} x^2}+V_0(x)$ with
 \[ V_{0}(x) = \left\{
   \begin{array}{l l}
     \frac{x^{2}}{2} & \quad \text{if $x>0$  }\\
     \infty & \quad \text{if $x \leq 0.$ }\\
   \end{array} \right.  \]
 The eigenvalues of $H_0$ take the form $E_n=2n+\frac{3}{2}$ with $n\in\mathbb{N}$ 
 and the corresponding eigenfunctions are
 \begin{equation}\nonumber 
  \psi_n(x)\propto\,x\,{\rm e}^{-x^2/2}\,_1\mbox{F}_1\bigg(-n;\frac{3}{2};x^2\bigg).
 \end{equation}
 $H_0$ can be seen as the harmonic oscillator in the reduced domain $(0,\infty)$ 
 and a null boundary condition at the origin, i.e., $\psi_n=0$ for $x=0$.
 Thus the eigenfunctions of $H_0$ are proportional to the odd eigenfunctions of the harmonic oscillator.
 There are also formal even eigenfunctions associated to $\mathcal{E}_n=2n+\frac{1}{2}$, $n\in \mathbb{N}$ given by
 \begin{equation}\label{chin}
  \chi_n(x)\propto\,{\rm e}^{-x^2/2}\,_1\mbox{F}_1\bigg(-n;\frac{1}{2};x^2\bigg),
 \end{equation}
 which, however, do not satisfy the boundary condition at $x=0$.

\subsection{First-order SUSY}
 In order to implement the first-order supersymmetric transformation (1-SUSY) one needs to define the {\it intertwining operator}
 \begin{equation}\nonumber
  A^+=\frac{1}{\sqrt{2}}\left(-\frac{d}{dx}+\alpha(x)\right),
 \end{equation}
 which differential order (one) defines precisely the order of the SUSY transformation. 
 Along with $A^+$, a second Hamiltonian $H_1=-\frac{1}{2}\frac{{\rm d}^2}{{\rm d} x^2}+V_1(x)$ is introduced,
 which is required to fulfill the {\it intertwining relation}
 \begin{equation}\nonumber
  H_1A^+=A^+H_0.
 \end{equation}
 $H_1$ and $H_0$ will be called {\it supersymmetric partners}. 
 Such an intertwining yields an immediate relation between the solutions of the stationary Schr\"odinger equation 
 $H_1\phi_n(x)=E_n\phi_n(x)$ and those of $H_0\psi_n(x)=E_n\psi_n(x)$, namely, 
 \begin{equation}\nonumber
  \phi_n(x)= C_nA^+\psi_n(x),
 \end{equation}
 where $C_n$ is a normalization constant. 

 For realizing the intertwining relation one needs only a {\it seed solution} $u(x)$ (or {\it transformation function})
 which, in general, is a solution of $H_0u(x)=\epsilon u(x)$ disregarding the boundary conditions and such that
 \begin{eqnarray}\nonumber
  A^+&=&\frac{1}{\sqrt{2}}\left[-\frac{d}{dx}+\left[\ln(u)\right]'\right],\\ \nonumber
  V_1&=&V_0-\left[\ln(u)\right]'',
 \end{eqnarray}
 where $\epsilon$ is the {\it factorization energy}. 
 Hereafter, if $f$ is a function of $x$ then $f'=\frac{df}{dx}$, $f''=\frac{d^2f}{dx^2}$, $\dots$ 
 shall be understood as common notation.

 The most general form of $u$ is 
 \begin{equation}\label{uepsilon}
  u(x,\epsilon)=e^{-x^2/2} \left[\,_1 F_1\left(\frac{1-2\epsilon}{4},\frac{1}{2},x^2\right)+2\nu \frac{\Gamma(\frac{3-2\epsilon}{4})}{\Gamma(\frac{1-2\epsilon}{4})}\,x\,_1 F_1\left(\frac{3-2\epsilon}{4},\frac{3}{2},x^2\right)\right], 
 \end{equation}
 which is a linear combination of an even and an odd seed solution. 
 Since one wishes to maintain control on the singularities appearing in the procedure,
 which are mainly a result of the terms proportional to $\ln(u)$ and its derivatives,
 hereafter only transformation functions with definite parity will be used.

 In \cite{fm14} it was found that for both choices of parity of $u(x)$ the first-order 
 SUSY partner Hamiltonians $H_1$ and $H_0$ are isospectral, up to a shift in the ground state energy.

\subsection{Second-order SUSY}
 If instead of the first-order operator $A^+$ one uses a second-order one, generally described as
 \begin{equation}\nonumber
  B^+=\frac{1}{2}\left(\frac{d^2}{dx^2}-\eta\frac{d}{dx}+\gamma\right),
 \end{equation}
 then one will implement a second-order supersymmetric transformation (2-SUSY).
 This time the intertwining relation takes the form 
 \begin{equation}\nonumber
  H_2B^+=B^+H_0,
 \end{equation}
 and the relation between solutions of $H_2\phi_n(x)=E_n\phi_n(x)$ and those of $H_0\psi_n(x)=E_n\psi_n(x)$ is now
 \begin{equation}\nonumber
 \phi_n(x)= D_nB^+\psi_n(x),
\end{equation}
 where $D_n$ is a normalization constant. 

 This second-order SUSY transformation is fixed now by two seed solutions of $H_0u_i=\epsilon_iu_i$, $i=1,2$, 
 for which without lost of generality $\epsilon_2<\epsilon_1$ is taken.
 The intertwining operator and the potential for the new Hamiltonian $H_2$ are given by
 \begin{eqnarray}\nonumber
  B^+&=&\frac{1}{2}\Bigg\{ \frac{d^2}{dx^2}-\left[\ln W\left(u_1,u_2\right)\right]'\frac{d}{dx}+\frac{1}{2}\left(\left[\ln W\left(u_1,u_2\right)\right]''+\left[\ln W\left(u_1,u_2\right)\right]'^2\right) -2V+\epsilon_1+\epsilon_2\Bigg\},\\ \nonumber
  V_2&=&V_0-\left[\ln W\left(u_1,u_2\right)\right]'',
 \end{eqnarray}
 where $W(f,g)$ is the Wronskian of the two functions $f$, $g$.

 The general form of the transformation functions $u_1$ and $u_2$ is a linear combination of 
 $\,xe^{-x^2/2}\,_1 F_1\left(\frac{3-2\epsilon_j}{4},\frac{3}{2},x^2\right)$ 
 and $\,e^{-x^2/2}\,_1 F_1\left(\frac{1-2\epsilon_j}{4},\frac{1}{2},x^2\right)$ (see Eq. (\ref{uepsilon}));
 however, this study will be focused on transformation functions with definite parity, for reasons previously stated.
 Indeed, four parity combinations for the pair $u_1(x)$, $u_2(x)$ are at hand \cite{fm14}:
 \begin{enumerate}
  \item If both $u_1(x)$ and $u_2(x)$ are taken to be odd, the transformation will be non-singular in the domain $(0,\infty)$ as long as
        $\epsilon_2<\epsilon_1\le\frac{3}{2}=E_0$ or $E_j=\frac{3+4j}{2}\le\epsilon_2<\epsilon_1\le\frac{3+4(j+1)}{2}=E_{j+1}$.
  \item When $u_1(x)$ is odd and $u_2(x)$ is even no singularities will be added to $H_2$ in $(0,\infty)$  if 
        ${\cal E}_j=\frac{1+4j}{2}\le\epsilon_2<\epsilon_1\le\frac{3+4j}{2}=E_j$.
  \item If $u_1(x)$ is chosen even and $u_2(x)$ odd, there will be no singularities other than the one at $x=0$ in the domain of $V_2$ as long as
        $\epsilon_2<\epsilon_1\le\frac{1}{2}={\cal E}_0$ or $E_j=\frac{3+4j}{2}\le\epsilon_2<\epsilon_1\le\frac{5+4j}{2}={\cal E}_{j+1}$. 
  \item When both $u_1(x)$ and $u_2(x)$ are taken to be even no new singularities are added to the new potential for $x>0$ if  
        $\epsilon_2<\epsilon_1\le\frac{1}{2}={\cal E}_0$ or ${\cal E}_j=\frac{1+4j}{2}\le\epsilon_2<\epsilon_1\le\frac{1+4(j+1)}{2} = {\cal E}_{j+1}$. 
 \end{enumerate}

 Using the second-order SUSY transformation, levels can be added or erased in the energy spectrum of $H_2$
 for particular choices of $\epsilon_1$ and $\epsilon_2$ but, in general, for the first three previous cases it turns out 
 that the eigenfunctions of the initial Hamiltonian transform into eigenfunctions of the new Hamiltonian $H_2$.
 On the other hand, for the last choice of parity combination, the eigenfunctions of the initial Hamiltonian transform 
 into non-physical solutions of the Schr\"odinger equation of $H_2$ and the non-physical solutions of the Schr\"odinger 
 equation of $H_0$ associated to $\mathcal{E}_n$ of Eq. (\ref{chin}) transform into the eigenfunctions of $H_2$.

\section{Painlev\'e IV equation}\label{PIV}
 Now consider the set of operators $\lbrace H,L^+,L^- \rbrace$ satisfying the following commutators 
 \begin{eqnarray}\nonumber
  &&[H,L^\pm]=\pm L^\pm,\\ \nonumber
  &&[L^-,L^+]=P_m(H), 
 \end{eqnarray}
 where $P_m(H)$ is a polynomial of degree $m$ in the Hamiltonian $H$ (of Schr\"odinger form) 
 and $L^\pm$ are $(m+1)$th-order differential operators. 
 Such algebraic structures are known as Polynomial Heisenberg Algebra (PHA) of order $m$,
 which can be recognized as deformations of the algebra of the harmonic oscillator.

 In particular, if the ladder operators $L^\pm$ are of third-order, then $m=2$ and the PHA is of second-order. 
 Thus, by factorizing $L^+=L_1^+L_2^+$ and $L^-=L_2^-L_1^-$, where $L_{1,2}^-=(L_{1,2}^+)^\dagger$ and
 $L_1^+=\frac{1}{\sqrt{2}}\left[-\frac{d}{dx}+f(x)\right]$, 
 $L_2^+=\frac{1}{2}\left[\frac{d^2}{dx^2}+g(x)\frac{d}{dx}+h(x)\right]$,
 the following set of equations is found: 
 \begin{eqnarray} \nonumber
  &f(x)=x+g(x),\\ \nonumber
  &h=\frac{g'}{2} - \frac{g^2}{2} - 2xg - x^2 + \varepsilon_2 + \varepsilon_3 - 2 \varepsilon_1 - 1, \\ \nonumber
  &V=\frac{x^2}{2} - \frac{g'}{2} + \frac{g^2}{2} + xg + \varepsilon_1 - \frac{1}{2},\\ \nonumber
  &\frac{d^2g}{dx^2}=\frac{1}{2g}\left(\frac{dg}{dx}\right)^2+\frac{3}{2}g^3+4xg^2+2(x^2-a)g+\frac{b}{g}. 
 \end{eqnarray}
 This last equation for the function $g(x)$ is recognized as the Painlev\'e IV (PIV) equation 
 with parameters $a=\varepsilon_2+\varepsilon_3-2\varepsilon_1-1$, $b=-2(\varepsilon_2-\varepsilon_3)^2$.
 In the standard approach one would solve the PIV equation in order to realize a second-order PHA; however, 
 it is possible to use a known realization of the second-order PHA, through supersymmetric partners 
 of the truncated oscillator, to obtain several solutions to this non-linear differential equation.

 A straightforward method to obtain such solutions is to analyze the operator product $L^+L^-$,
 \begin{equation}\nonumber
  L^+L^-=\left(H-\varepsilon_1\right)\left(H-\varepsilon_2\right)\left(H-\varepsilon_3\right),
 \end{equation}
 which indicates that there are three extremal states $\phi_{\varepsilon_1}$, $\phi_{\varepsilon_2}$ and $\phi_{\varepsilon_3}$, 
 with corresponding eigenvalues $\varepsilon_1$, $\varepsilon_2$ and $\varepsilon_3$ 
 respectively which are simultaneously annihilated by $L^-$.

 One of these extremal states, say $\phi$, satisfies the equation
 \begin{equation} \nonumber
  L_1^-\phi=\frac{1}{\sqrt{2}}\left[\frac{d}{dx}+f(x)\right]\phi=0,
 \end{equation}
 where $f(x)=x+g(x)$. This can be solved for $g(x)$ as 
 \begin{equation}\nonumber
  g(x)=-x-[\ln\phi]',
 \end{equation}
 which is a solution to the PIV equation with parameters 
 $a=\varepsilon_2+\varepsilon_3-2\varepsilon_1-1$, $b=-2(\varepsilon_2-\varepsilon_3)^2$.
 One can use any extremal state as $\phi$ to calculate $g(x)$, 
 thus there are three solutions $g_i(x)=-x-[\ln\phi_{\varepsilon_i}]'$, $i=1,2,3$ 
 corresponding to the three different choices of $\phi$. 

 If $H=H_1$ one can define natural ladder operators $L^\pm=A^+a^\pm A$, 
 where $a^\pm$ are the usual first-order ladder operators of the harmonic oscillator. 
 Therefore, $L^\pm$ are differential operators of third-order and $\lbrace H,L^+,L^- \rbrace$ realize a second-order PHA. 
 Note, however, that operators $L^\pm$ move, in general, a solution of the Schr\"odinger equation from a 
 physical state $E_n$ to a non-physical solution associated to $\mathcal{E}_n$ and vice versa. 

 We conclude that a family of realizations of second-order PHA's were found 
 through first-order supersymmetric transformations on the truncated oscillator and thus 
 several solutions of the PIV equation can be found departing from the following extremal states of $L^-$
 \begin{equation}\nonumber
  \phi_{\varepsilon_1}\propto \frac{1}{u}, \qquad \phi_{\varepsilon_2}\propto A^+a^+u,\qquad  \phi_{\varepsilon_3}\propto A^+e^{\frac{-x^2}{2}},
 \end{equation}
 with corresponding (formal) eigenvalues $\epsilon$, $\epsilon+1$, $\frac{1}{2}$.

 When $u(x)$ is odd the first identification of extremal states gives a solution to the PIV equation of the form
 \begin{equation}\nonumber
  g_1=\frac{1}{x}-2x+\left(1-\frac23\epsilon\right) x \, \frac{\,_1F_1(\frac{7-2\epsilon}{4};\frac{5}{2};x^2)}{\,_1F_1(\frac{3-2\epsilon}{4};\frac{3}{2};x^2)},
 \end{equation}
 while if $u(x)$ is even the solution to the PIV equation obtained for the first identifications is
 \begin{equation}\nonumber
  g_1=-2x+(1-2\epsilon)\, x \, \frac{\,_1F_1(\frac{5-2\epsilon}{4};\frac{3}{2};x^2)}{\,_1F_1(\frac{1-2\epsilon}{4};\frac{1}{2};x^2)}.
 \end{equation}
 In both cases the parameters of the PIV equation are 
 $a_1=-\epsilon+\frac{1}{2}$, $b_1=-2\left(\epsilon + \frac{1}{2}\right)^2$.

 The solutions obtained through the other two permutations of the extremal states can be written in terms of $g_1(x)$:
 \begin{eqnarray}\nonumber
  &g_2=-g_1-2x-2 \left[\frac{x + (2\epsilon - x^2)(g_1+x) + (g_1+x)^3}{x^2-2\epsilon-1-(g_1+x)^2} \right], \\ \nonumber
  &g_3=-\frac{g_1'+2}{g_1 + 2x}.
 \end{eqnarray}
 The parameters of the PIV equation are now $a_2=-\epsilon-\frac{5}{2}$, $b_2=-2\left(\epsilon-\frac{1}{2}\right)^2$, 
 and $a_3=2\epsilon-1$, $b_3=-2$ respectively.

 On the other hand, for $H=H_2$ one can define natural ladder operators $L^\pm=B^+a^\pm B$
 which are of fifth-order; thus, one needs to use the reduction theorem described in \cite{bf11a} 
 which asserts that when $u_2(x)=a^-u_1(x)$, and $\epsilon_2=\epsilon_1-1$, the Hamiltonian $H_2$ 
 has a set of third-order ladder operators $l^\pm$ such that
 \begin{equation}\nonumber
 l^+l^-=(H_2-\epsilon_1+1)(H_2-\epsilon_1-1)(H_2-1/2). 
 \end{equation}

 Once again the set $\lbrace H_2,l^+,l^- \rbrace$ realizes the second-order PHA and the sought extremal states are now
 \begin{equation}\nonumber
  \phi_{\varepsilon_1}\propto \frac{u_1}{W[u_1,u_2]}, \qquad  \phi_{\varepsilon_2}\propto B^+a^+u_1, \qquad  \phi_{\varepsilon_3}\propto B^+e^{\frac{-x^2}{2}},
 \end{equation}
 which correspond to the (formal) eigenvalues $\epsilon_1-1$, $\epsilon_1+1$, $\frac{1}{2}$. 

 A similar procedure as for $H_1$ gives the PIV solutions $G_i(x)$ in terms of $\alpha=\frac{u_1'}{u_1}$ in the following way
 \begin{eqnarray}\nonumber
  G_{1} & =&  - x - \alpha + 2\left[\frac{x+\alpha}{x^2+1-2\epsilon_1-\alpha^2}\right], \\ \nonumber
  G_{2} & =&   G_1+\frac{2\alpha^2-2x^2+2(2\epsilon_1+1)}{\alpha-G_1-x}, \\ \nonumber
  G_{3} & =&  \frac{(x+\alpha)G_1^2+\left[2\epsilon_1-1+(x+\alpha)^2\right]G_1+(2\epsilon_1-3)(x+\alpha)}{(x+\alpha)^2+(x+\alpha)G_1+2\epsilon_1-1},
 \end{eqnarray}
 associated to the parameters $a_1=-\epsilon_1+\frac{5}{2}$ and $b_1=-2\left(\epsilon_1+\frac{1}{2}\right)^2$, 
 $a_2=-\epsilon_1-\frac{7}{2}$ and $b_2=-2\left(\epsilon_1-\frac{3}{2}\right)^2$, 
 $a_3=2(\epsilon_1-1)$ and $b_3=-8$, respectively.

\section{Painlev\'e V equation}\label{PV}
 In the last section, the second-order PHA's were connected to the PIV equation. 
 Is there something similar for third-order PHA's?
 Consider a set $\lbrace H,L^+,L^- \rbrace$, where $L^\pm$ are fourth-order ladder operators such that
 \begin{eqnarray}\nonumber
  &&[H,L^\pm]=\pm 2L^\pm,\\ \nonumber
  &&[L^-,L^+]=P_3(H), 
 \end{eqnarray}
 $P_3(H)$ being a polynomial of third degree in the Hamiltonian $H$, i.e., we are dealing with a third-order PHA.

 To realize such an algebraic structure in terms of differential operators, choose $L^+=L_1^+L_2^+$ and $L^-=L_2^-L_1^-$,
 where $L_1^+=\frac{1}{2}\left[\frac{d^2}{dx^2}+g_1(x)\frac{d}{dx}+h_1(x)\right]$, 
 $L_2^+=\frac{1}{2}\left[\frac{d^2}{dx^2}+g(x)\frac{d}{dx}+h(x)\right]$ and $L_{1,2}^-=(L_{1,2}^+)^\dagger$;
 one finds now this other set of equations: 
 \begin{eqnarray} \nonumber
  &g_1(x)=-2x-g(x),\\ \nonumber
  &g(x)=\frac{2x}{w-1},\\  \nonumber
  &\frac{d^2w}{dz^2}=\left(\frac{1}{2w}+\frac{1}{w-1}\right)\left(\frac{dw}{dz}\right)^2-\frac{1}{z}\frac{dw}{dz}+\frac{(w-1)^2}{z^2}(aw+\frac{b}{w})+c\frac{w}{z}+d\frac{w(w+1)}{w-1}.
 \end{eqnarray}
 The last equation is recognized as the Painlev\'e V (PV) equation with parameters 
 $a=\frac{(\varepsilon_1-\varepsilon_2)^2}{8}$, $b=-\frac{(\varepsilon_3-\varepsilon_4)^2}{8}$, 
 $c=\frac{\varepsilon_1+\varepsilon_2-\varepsilon_3-\varepsilon_4}{4}-\frac{1}{2}$ and 
 $d=-\frac{1}{8}$, where $z=2x^2$ has been used.

 The solution $w(z)$ of the PV equation characterizes completely the third-order PHA described above. 
 Conversely, one can use known realizations of the third-order PHA to obtain solutions to the PV equation. 
 In this section one uses the supersymmetric partners of the truncated oscillator to build such realizations.
 For this purpose, let us study the product $L^+L^-$, factorized as 
 \begin{equation}\nonumber
  L^+L^-=\left(H-\varepsilon_1\right)\left(H-\varepsilon_2\right)\left(H-\varepsilon_3\right)\left(H-\varepsilon_4\right).
 \end{equation}
 This indicates that $H$ has four extremal states $\phi_{\varepsilon_1}$, $\phi_{\varepsilon_2}$, $\phi_{\varepsilon_3}$ and $\phi_{\varepsilon_4}$ 
 corresponding to the eigenvalues $\varepsilon_1$, $\varepsilon_2$, $\varepsilon_3$ and $\varepsilon_4$ respectively which are also annihilated by $L^-$.

 It can be shown now that
 \begin{equation}\nonumber
  w(z)=1+\frac{\sqrt{2z}}{g(\sqrt{\frac{z}{2}})}
 \end{equation}
 is a solution of the PV equation as long as 
 $g(x)=- x-\left(\text{ln}\left[W(\phi_{\varepsilon_3},\phi_{\varepsilon_4})\right]\right)'$ and $x=\sqrt{\frac{z}{2}}$.
 Among the four extremal states the choice of $\phi_{\varepsilon_3}$, $\phi_{\varepsilon_4}$ is arbitrary, 
 so one has six identifications for said pair.

 For $H=H_1$, along with the third-order ladder operators of the last section one can define fourth-order ones 
 $L^\pm=A^+(a^\pm)^2 A$, where $a^\pm$ are again the usual first-order ladder operators of the harmonic oscillator. 
 Note that this time the ladder operators move among physical levels in the energy spectrum, i.e., 
 their steps have size two. 

 Thus, solutions to the PV equation are at hand through the procedure detailed above, with the extremal states of $L^-$ being now
 \begin{equation}\nonumber
  \phi_{\varepsilon_1}\propto \frac{1}{u}, \qquad \phi_{\varepsilon_2}\propto A^+(a^+)^2u, \qquad \phi_{\varepsilon_3}\propto A^+\chi_0, \qquad  \phi_{\varepsilon_4}\propto A^+\psi_0, 
 \end{equation}
 corresponding to the (formal) eigenvalues $\epsilon$, $\epsilon+2$, $\frac{1}{2}$, $\frac{3}{2}$, respectively.
 Then, for each identification $\lbrace\varepsilon_i\rbrace=\lbrace \epsilon,\epsilon+2,\frac{1}{2},\frac{3}{2}\rbrace$,
 $i=1,2,3,4$, the solution $w(z)$ to the PV equation can be straightforwardly calculated.
 Thus, we get the following solutions ($\alpha=\frac{u'}{u}$):

 \begin{enumerate}[(a)]
  \item For the choice $\varepsilon_1=\frac{1}{2}$, \,$\varepsilon_2=\frac{3}{2}$, \,$\varepsilon_3=\epsilon$, \,$\varepsilon_4=\epsilon+2$, 
        the function
        \begin{equation}\nonumber
         w_{1a}(z)=1+\frac{2\sqrt{2z}(1+2\epsilon -z+\sqrt{2z}\alpha)}{\sqrt{2z}(2+z)-4(1+z)\alpha+2\sqrt{2z}\alpha^2}
        \end{equation}
        solves the PV equation with parameters $a=\frac{1}{8}$, $b=-\frac{1}{2}$, $c=-\frac{1}{2}(\epsilon+1)$.
  \item For the choice $\varepsilon_1=\epsilon$, \,$\varepsilon_2=\frac{3}{2}$, \,$\varepsilon_3=\frac{1}{2}$, \,$\varepsilon_4=\epsilon+2$,
        the function 
        \begin{equation}\nonumber
         w_{1b}(z)=\frac{\left(2 \alpha +\sqrt{2z}\right) \left(4 \alpha -4\epsilon \sqrt{2z}+\sqrt{2} z^{3/2}-2 \sqrt{2} \alpha ^2 \sqrt{z}
         -4 \sqrt{2z}\right)}{\left(2 \alpha-\sqrt{2z} \right) \left(4 \alpha -4\epsilon \sqrt{2z}+\sqrt{2} z^{3/2}-2 \sqrt{2} \alpha ^2 \sqrt{z}+4
         \sqrt{2z}\right)}     
        \end{equation}
        solves the PV equation with parameters $a=\frac{1}{8}\left(\epsilon-\frac{3}{2}\right)^2$, $b=-\frac{1}{8}\left(\epsilon+\frac{3}{2}\right)^2$, $c=-\frac{3}{4}$.   
  \item For the choice $\varepsilon_1=\frac{1}{2}$, \,$\varepsilon_2=\epsilon$, \,$\varepsilon_3=\frac{3}{2}$, \,$\varepsilon_4=\epsilon+2$,
        the function 
        \begin{equation}\nonumber
         w_{1c}(z)=1+\frac{\sqrt{2z} \left(8 \epsilon z-2 z^2+4 \alpha ^2 z+4 \sqrt{2z} \alpha-16\right)}{8 \alpha -2 \sqrt{2} z^{3/2} \left(\alpha ^2+2
         \epsilon-3\right)+4 \alpha  z \left(\alpha ^2+2 \epsilon-1\right)-8 \sqrt{2z} \left(\alpha ^2+\epsilon-1\right)+\sqrt{2} z^{5/2}-2 \alpha  z^2}
        \end{equation}
        solves the PV equation with parameters $a=\frac{1}{8}\left(\epsilon-\frac{1}{2}\right)^2$, $b=-\frac{1}{8}\left(\epsilon+\frac{1}{2}\right)^2$, $c=-\frac{5}{4}$. 
  \item For the choice $\varepsilon_1=\frac{1}{2}$, \,$\varepsilon_2=\epsilon+2$, \,$\varepsilon_3=\epsilon$, \,$\varepsilon_4=\frac{3}{2}$,
        the function 
        \begin{equation}\nonumber
         w_{1d}(z)=\frac{2 \sqrt{2}-2 \alpha  \sqrt{z}-\sqrt{2} z}{2 \sqrt{2}-2 \alpha  \sqrt{z}+\sqrt{2} z}
        \end{equation}
        solves the PV equation with parameters $a=\frac{1}{8}\left(\epsilon+\frac{3}{2}\right)^2$, $b=-\frac{1}{8}\left(\epsilon-\frac{3}{2}\right)^2$, $c=-\frac{1}{4}$. 
  \item For the choice $\varepsilon_1=\epsilon+2$, \,$\varepsilon_2=\frac{3}{2}$, \,$\varepsilon_3=\epsilon$, \,$\varepsilon_4=\frac{1}{2}$,
        the function 
        \begin{equation}\nonumber
         w_{1e}(z)=\frac{2 \alpha +\sqrt{2z}}{2 \alpha-\sqrt{2z} }
        \end{equation}
        solves the PV equation with parameters $a=\frac{1}{8}\left(\epsilon+\frac{1}{2}\right)^2$, $b=-\frac{1}{8}\left(\epsilon-\frac{1}{2}\right)^2$, $c=\frac{1}{4}$.
  \item For the choice $\varepsilon_1=\epsilon$, \,$\varepsilon_2=\epsilon+2$, \,$\varepsilon_3=\frac{1}{2}$, \,$\varepsilon_4=\frac{3}{2}$,
        the function 
        \begin{equation}\nonumber
         w_{1f}(z)=-\frac{-4 \alpha +\sqrt{2} z^{3/2}+2 \sqrt{2} \left(\alpha ^2-1\right) \sqrt{z}+4 \alpha  z}{4 \alpha -2 \sqrt{2} \sqrt{z} \left(\alpha ^2+2
	 \epsilon-2\right)+\sqrt{2} z^{3/2}}
        \end{equation}
        solves the PV equation with parameters $a=\frac{1}{2}$, $b=-\frac{1}{8}$, $c=\frac{1}{2}(\epsilon-1)$. 
 \end{enumerate}

 On the other hand, for $H=H_2$ one can define natural ladder operators through $L^\pm=B^+(a^\pm)^2B$,
 but these are of sixth-order, thus one needs again a reduction theorem. 
 Such a theorem is a slight generalization of the one in \cite{bf11a}, but for energy spectra of arbitrary constant spacing.
 In the present case, where the spacing is two, it asserts that when $u_2=(a^-)^2u_1$ and $\epsilon_2=\epsilon_1-2$, 
 there exists also a pair of fourth-order ladder operators $l^\pm$ for $H_2$ such that
 \begin{equation}\nonumber
  l^+l^-=\left(H_2-\epsilon_1+2\right)\left(H_2-\epsilon_1-2\right)\left(H_2-\frac{1}{2}\right)\left(H_2-\frac{3}{2}\right).
 \end{equation}

 Therefore, the extremal states to be used in the procedure are
 \begin{equation}\nonumber
  \phi_{\varepsilon_1}\propto \frac{u_1}{W(u_1,u_2)}, \qquad \phi_{\varepsilon_2}\propto B^+(a^+)^2u_1,\qquad \phi_{\varepsilon_3}\propto B^+\chi_0,\qquad \phi_{\varepsilon_4}\propto B^+\psi_0
 \end{equation}
 associated to the (formal) eigenvalues $\epsilon_1-2$, $\epsilon_1+2$, $\frac{1}{2}$, $\frac{3}{2}$, respectively.

 As before, for every identification $\lbrace\varepsilon_i\rbrace=\lbrace\frac{1}{2},\frac{3}{2},\epsilon_1-2,\epsilon_1+2\rbrace$, $i=1,2,3,4$, 
 one calculates first $g(x)=-2x-\left(\text{ln}\left[W(\phi_3,\phi_4)\right]\right)'$ to later obtain
 \begin{equation}\nonumber
  w(z)=1+\frac{\sqrt{2z}}{g(\sqrt{\frac{z}{2}})},
 \end{equation}
 which is a solution of the PV equation with parameters $a=\frac{(\varepsilon_1-\varepsilon_2)^2}{8}$, 
 $b=-\frac{(\varepsilon_3-\varepsilon_4)^2}{8}$, $c=\frac{\varepsilon_1+\varepsilon_2-\varepsilon_3-\varepsilon_4}{4}-\frac{1}{2}$
 and $d=-\frac{1}{8}$. Note that these parameters, and therefore the PV equations they characterize, are symmetric under the 
 changes $\varepsilon_1\leftrightarrow\varepsilon_2$ and $\varepsilon_3\leftrightarrow\varepsilon_4$; 
 thus, one has the following six distinct identifications.

 \begin{enumerate}[(a)]
  \item $\varepsilon_1=\frac{1}{2}$, \,$\varepsilon_2=\frac{3}{2}$, \,$\varepsilon_3=\epsilon_1-2$, \,$\varepsilon_4=\epsilon_1+2$
        with parameters $a=\frac{1}{8}$, $b=-2$, $c=-\frac{\epsilon_1}{2}$. 
  \item $\varepsilon_1=\epsilon_1-2$, \,$\varepsilon_2=\frac{3}{2}$, \,$\varepsilon_3=\frac{1}{2}$, \,$\varepsilon_4=\epsilon_1+2$
        with parameters $a=\frac{1}{8}\left(\epsilon_1-\frac{7}{2}\right)^2$, $b=-\frac{1}{8}\left(\epsilon_1+\frac{3}{2}\right)^2$, $c=-\frac{5}{4}$. 
  \item $\varepsilon_1=\frac{1}{2}$, \,$\varepsilon_2=\epsilon_1-2$, \,$\varepsilon_3=\frac{3}{2}$, \,$\varepsilon_4=\epsilon_1+2$
        with parameters $a=\frac{1}{8}\left(\epsilon_1-\frac{5}{2}\right)^2$, $b=-\frac{1}{8}\left(\epsilon_1+\frac{1}{2}\right)^2$, $c=-\frac{7}{4}$.  
  \item $\varepsilon_1=\frac{1}{2}$, \,$\varepsilon_2=\epsilon_1+2$, \,$\varepsilon_3=\epsilon_1-2$, \,$\varepsilon_4=\frac{3}{2}$
        with parameters $a=\frac{1}{8}\left(\epsilon_1+\frac{3}{2}\right)^2$, $b=-\frac{1}{8}\left(\epsilon_1-\frac{7}{2}\right)^2$, $c=\frac{1}{4}$. 
  \item $\varepsilon_1=\epsilon_1+2$, \,$\varepsilon_2=\frac{3}{2}$, \,$\varepsilon_3=\epsilon_1-2$, \,$\varepsilon_4=\frac{1}{2}$
        with parameters $a=\frac{1}{8}\left(\epsilon_1+\frac{1}{2}\right)^2$, $b=-\frac{1}{8}\left(\epsilon_1-\frac{5}{2}\right)^2$, $c=\frac{3}{4}$. 
  \item $\varepsilon_1=\epsilon_1-2$, \,$\varepsilon_2=\epsilon_1+2$, \,$\varepsilon_3=\frac{1}{2}$, \,$\varepsilon_4=\frac{3}{2}$
        with parameters $a=2$, $b=-\frac{1}{8}$, $c=\frac{1}{2}(\epsilon_1-2)$. 
  \end{enumerate}

 As an example of a solution of the PV equation obtained through this method, 
 let us fix a factorization energy $\epsilon_1=-\frac{5}{2}$ and then let us take an even transformation function $u(x)$ to get
 \begin{equation}\nonumber
  w_{2a}(z)=-\frac{4 (z+1)}{z^2-2 z-1},
 \end{equation}
 for the PV parameters $a=\frac{1}{8}$, $b=-2$, $c=\frac{5}{4}$, $d=-\frac{1}{8}$ in the case (a); 
 \begin{equation}\nonumber
  w_{2d}(z)=\frac{z^3+13 z^2+65 z+105}{2 z^2+20 z+30}
 \end{equation}
 for the PV parameters $a=\frac{1}{2}$, $b=-\frac{49}{8}$, $c=\frac{1}{4}$, $d=-\frac{1}{8}$ in the case (d) with an 
 odd transformation function $u(x)$ associated to $\epsilon_1=-\frac{7}{2}$; 
 or take the case (f) with an odd transformation function $u(x)$ for $\epsilon_1=-\frac{3}{2}$ to get
 \begin{equation}\nonumber
  w_{2f}(z)=-\frac{z^2+2 z+3}{4 (z+3)}
 \end{equation}
 associated to the PV parameters $a=2$, $b=-\frac{1}{8}$, $c=-\frac{7}{4}$, $d=-\frac{1}{8}$.

\section{B\"acklund transformations}
 Several solutions of the PIV and PV equations have been obtained in the previous sections using a method
 that exploits the connection between PHA of second and third-order, realized by supersymmetric partners of
 the truncated oscillator, and these two non-linear second-order differential equations.

 In each case the solutions were obtained through a simple method which, however does not show clearly a possible relation 
 among them. This section shall provide an insight on this relation in the language of B\"acklund transformations. 

 A B\"acklund transformation (BT) between two partial differential equations 
 \begin{equation}\nonumber
  D(u;x,t)=0,\qquad E(V;X,T)=0
 \end{equation}
 is generally defined as a set of relations involving $\{x,t,u(x,t)\}$, $\{X,T,V(X,T)\}$ and derivatives of $u$ and $V$ 
 such that \cite{as81}:
 \begin{enumerate}
  \item the BT is integrable for $V$ if and only if $D(u)=0$;
  \item the BT is integrable for $u$ if and only if $E(V)=0$;
  \item given $u$ such that $D(u)=0$, the BT defines $V$ up to a finite number of constants, and $E(V)=0$;
  \item given $V$ such that $E(V)=0$, the BT defines $u$ up to a finite number of constants, and $D(u)=0$.
 \end{enumerate}
 (Remember that $v_x=f(x,t)$ and $v_t=g(x,t)$ are integrable for $v$ iff $v_{xt}=v_{tx}$; i.e., they must be compatible).

 Alternatively, as described informally in \cite{bch95}, 
 a B\"acklund transformation relates one solution of a given equation
 either to another solution of the same equation, 
 possibly with different values of the parameters, or to a solution of another equation.

 In the following subsections we will find the B\"acklund transformations 
 relating the previously obtained solutions of the PIV and PV equations separately.
 For some transformations the calculations are shown explicitly; however, for others
 only the final results will be depicted, due to the length of the expressions involved.

\subsection{Painlev\'e IV}
In the language of \cite{bch95}, the chain of B\"acklund transformation relating the solutions of the PIV equation 
obtained in Section \ref{PIV} are schematically shown as
\begin{eqnarray} \nonumber
 1-SUSY \qquad\qquad\qquad &\qquad& \qquad\qquad\qquad 2-SUSY\\ \nonumber
 \overbrace{g_1 \qquad\to\qquad g_2 \qquad\to\qquad g_3} \qquad&\to&\qquad \overbrace{G_1 \qquad\to\qquad G_3 \qquad\to\qquad G_2}\\ \nonumber
  W^{\ddag+}W^{\dagger+}\qquad\qquad W^{\ddag-} \qquad\qquad &\tilde{W}^+& \qquad\qquad\quad W^{\ddag-} \qquad\qquad W^{\ddag+}\nonumber
\end{eqnarray}
where 
\begin{eqnarray}\label{wtilde}
 \tilde{W}^+\left[g(x;a,b)\right] &:=& \tilde{g}(x;\tilde{a},\tilde{b})=\frac{g'-g^2-2xg-\sqrt{-2b}}{2g},\\ \nonumber
   \tilde{a}&=& \frac{1}{4}\left(1-2a+3\sqrt{-2b}\right),\\ \nonumber
   \tilde{b}&=& -\frac{1}{2}\left(1+a+\frac{1}{2}\sqrt{-2b} \right)^2 ,\\ \label{wdagger}
 W^{\dagger+}\left[g(x;a,b)\right] &:=& g^\dagger(x;a^\dagger,b^\dagger)= g+\frac{2(1-a-\frac{1}{2}\sqrt{-2b})g}{g'+\sqrt{-2b}+2xg+g^2},\\ \nonumber
   a^\dagger&=&\frac{3}{2}-\frac{a}{2}-\frac{3}{4}\sqrt{-2b} ,\\ \nonumber
   b^\dagger&=& -\frac{1}{2}\left(1-a+\frac{1}{2}\sqrt{-2b} \right)^2 ,\\ \label{wddag}
 W^{\ddag\pm}\left[g(x;a,b)\right] &:=& g^\ddag(x;a^\ddag,b^\ddag)= g+\frac{2(1+a\pm\frac{1}{2}\sqrt{-2b})g}{g'\mp\sqrt{-2b}-2xg-g^2} ,\\ \nonumber 
   a^\ddag&=&-\frac{3}{2}-\frac{a}{2}\mp\frac{3}{4}\sqrt{-2b} ,\\ \nonumber
   b^\ddag&=& -\frac{1}{2}\left(-1-a\pm\frac{1}{2}\sqrt{-2b} \right)^2.
\end{eqnarray}

First of all this diagram is valid for the pair of parameters $a$, $b$ of each equation, but, 
Is this also true for the solutions? The answer is positive, and this comes about from the following reasoning.

From the Schr\"odinger equation, rewritten as $u''=(x^2-2\epsilon)u$, and the expression $g_1=-x+\alpha$ one gets
$g_1'=-1+\frac{u''}{u}-\alpha^2=-1+x^2-2\epsilon-(x+g_1)^2$. Now, use the expression for $g_2$ in terms of $g_1$,
\begin{equation}\nonumber
 g_2=-g_1-2x-2\left(\frac{x+(2\epsilon-x^2)(g_1+x)+(g_1+x)^3}{x^2-2\epsilon-1-(g_1+x)^2} \right),
\end{equation}
to rewrite $g_2$ as
\begin{eqnarray}\nonumber
 g_2&=&g_1-2g_1-2x-2\left(\frac{x+(2\epsilon-x^2)(g_1+x)+(g_1+x)^3}{g_1'} \right)\\ \nonumber
  &=&g_1+2\left(\frac{-g_1'(g_1+x)-x-(2\epsilon-x^2)(g_1+x)-(g_1+x)^3}{g_1'} \right)\\ \nonumber
  &=&g_1+2\left(\frac{-g_1'(g_1+x)-x+(1-1-2\epsilon+x^2-(g_1+x)^2)(g_1+x)}{g_1'} \right)\\ \nonumber
  &=&g_1+2\left(\frac{-g_1'(g_1+x)-x+(1+g_1')(g_1+x)}{g_1'} \right)\\ \nonumber
  &=&g_1+2\left(\frac{g_1}{g_1'} \right).
\end{eqnarray}
This last equation can be expanded into
\begin{eqnarray}\nonumber
 g_2&=&g_1+\left(\frac{4g_1}{2g_1'} \right)\\ \nonumber
  &=&g_1+\left(\frac{4g_1}{g_1'+x^2-2\epsilon-1-(x+g_1)^2} \right)\\ \nonumber
  &=&g_1+\left(\frac{4g_1}{g_1'-2\epsilon-1-2xg_1-g_1^2} \right).
\end{eqnarray}
This expression coincides with the result of the third transformation in table 4a in \cite{bch95} applied to $g_1$:
\begin{equation}\nonumber
 W^{\ddag+}W^{\dagger+}\left[g_1\right]=g_1+\left(\frac{4g_1}{g_1'-2\epsilon-1-2xg_1-g_1^2} \right),
\end{equation}
which establishes the validity of the first transformation in the diagram.

Now, from the results $g_2=g_1+2\frac{g_1}{g_1'} $ and $g_1'=-2\epsilon-1-2xg_1-g_1^2$ one can obtain 
$g_2'=g_1'+2-2\frac{g_1g_1''}{g_1'^2}$, where $g_1''=2\left(x-(x+g_1)(1+g_1')\right)$, such that
\begin{eqnarray}\nonumber
 g_1''&=&2\left( g_1^3+x+3xg_1^2+2x \epsilon +2g_1(x^2+ \epsilon ) \right), \\ \nonumber
 g_2'&=&1-2\epsilon-g_1^2-2xg_1-4g_1\frac{g_1^3+x+3xg_1^2+2x^2g_1+2\epsilon(x+g_1)}{(1+2\epsilon+g_1^2+2xg_1)^2}.
\end{eqnarray}
Then one can write
\begin{equation}\nonumber
 g_2'+2\epsilon-1-2xg_2-g_2^2=-2g_1(2x+g_1).
\end{equation}

Using this last equation to simplify the transformation 
$W^{\ddag-}\left[g_2\right]=g_2-\frac{(4\epsilon+2)g_2}{g_2'+2\epsilon-1-2xg_2-g_2^2}$, it turns out that
\begin{eqnarray}\nonumber
 W^{\ddag-}\left[g_2\right]&=&g_2+\frac{(2\epsilon+1)g_2}{g_1(2x+g_1)}\\ \nonumber
  &=&\frac{(2+g_1')(1+2\epsilon+g_1^2+2xg_1)}{g_1'(2x+g_1)}\\ \nonumber
  &=&-\frac{2+g_1'}{2x+g_1}.
\end{eqnarray}
The last is exactly the expression for $g_3$ in terms of $g_1$ found in Section \ref{PIV},
which proves the second transformation in the diagram.

To establish the third transformation differentiate first $g_3$ 
\begin{equation}\nonumber
 g_3'=\frac{(2+g_1')^2+2(g_1+2x)(xg_1'+g_1g_1'+g_1)}{(g_1+2x)^2}. 
\end{equation}
Thus
\begin{equation}
 g_3'-g_3^2-2xg_3-2=2g_1',
\end{equation}
which is used to calculate the following transformation
\begin{eqnarray}\nonumber
 \tilde{W}^+\left[g_3\right]&=&\frac{g_3'-g_3^2-2xg_3-2}{2g_3}\\ \nonumber
 &=&-\frac{g_1'(g_1+2x)}{2+g_1'}=-x-\alpha+2\left(\frac{\alpha+x}{2+g_1'}\right).
\end{eqnarray}
Now, since $g_1=\alpha-x$ then $g_1'=-1-2\epsilon-2xg_1-g_1^2=x^2-1-2\epsilon-\alpha^2$ and we get
\begin{equation}\nonumber
 \tilde{W}^+\left[g_3\right]=-x-\alpha+2\left(\frac{\alpha+x}{x^2+1-2\epsilon-\alpha^2}\right).
\end{equation}
Therefore, one can conclude that $G_1=\tilde{W}^+\left[g_3\right]$.

Next, from $G_1=-x-\alpha+2\frac{x+\alpha}{x^2+1-2\epsilon-\alpha^2}$ and $\alpha'=x^2-2\epsilon-\alpha^2$,
some calculations show that
\begin{eqnarray}\nonumber
 G_1'&=&-1-\alpha'+2\frac{(x^2+1-2\epsilon-\alpha^2)(\alpha'+1)-(x+\alpha)(2x-2\alpha\alpha')}{(x^2+1-2\epsilon-\alpha^2)^2}\\ \label{G_1'}
  &=&-1+\alpha^2+2\epsilon-x^2+2\frac{(1-\alpha^2-2\epsilon+x^2)^2-2(\alpha+x)(\alpha^3+2\epsilon\alpha+x-x^2\alpha)}{(x^2+1-2\epsilon-\alpha^2)^2}.
\end{eqnarray}
Then transforming $G_1$ with $W^{\ddag-}$ one gets
\begin{eqnarray}\nonumber
W^{\ddag-}\left[G_1\right]&=&G_1+\frac{(6-4\epsilon)G_1}{G_1'+1+2\epsilon-2xG_1-G_1^2}\\ \nonumber
 &=&-x-\alpha+2\frac{x+\alpha}{x^2+1-2\epsilon-\alpha^2}+\frac{(2\epsilon-3)(\alpha+x)(\alpha^2+2\epsilon-x^2-1)}{1+4\epsilon^2+\alpha^2(2\epsilon-3)-4x\alpha-x^2-2\epsilon(2+x^2)}\\ \nonumber
 &=&\frac{4(\alpha+x)}{(\alpha^2+2\epsilon-x^2-1)}\\\nonumber
 &&\times\frac{\left(-1-4\epsilon^2+x\alpha^3-x^2-x^4+4\epsilon(1+x^2)+\alpha^2(2-2\epsilon+x^2)+\alpha(x+2x\epsilon-x^3)\right)}
{\left(1+4\epsilon^2+\alpha^2(2\epsilon-3)-4x\alpha-x^2-2\epsilon(x^2+2)\right)}.
\end{eqnarray}

On the other hand, if one substitutes $G_1=-x-\alpha+2\frac{x+\alpha}{x^2+1-2\epsilon-\alpha^2}$ into $G_3$ one gets
\begin{eqnarray} \nonumber
 G_3&=&\frac{(x+\alpha)G_1^2+\left(2\epsilon-1+(x+\alpha)^2\right)G_1+(2\epsilon-3)(x+\alpha)}{(x+\alpha)^2+(x+\alpha)G_1+2\epsilon-1}\\\nonumber
  &=&\frac{4(\alpha+x)}{(\alpha^2+2\epsilon-x^2-1)}\\\label{G3}
 &&\times\frac{\left(-1-4\epsilon^2+x\alpha^3-x^2-x^4+4\epsilon(1+x^2)+\alpha^2(2-2\epsilon+x^2)+\alpha(x+2x\epsilon-x^3)\right)}
{\left(1+4\epsilon^2+\alpha^2(2\epsilon-3)-4x\alpha-x^2-2\epsilon(x^2+2)\right)}.
\end{eqnarray}
Therefore $G_3=W^{\ddag-}\left[G_1\right]$.

Finally a similar treatment to the previous transformations shows that $G_2=W^{\ddag+}\left[G_3\right]$.
One starts with the expression for $W^{\ddag+}\left[G_3\right]$ as follows
\begin{equation}\nonumber
 W^{\ddag+}\left[G_3\right]=G_3+\frac{(4\epsilon+2)G_3}{G'_3+4-2xG_3-G_3^2}
\end{equation}
where $G_3$ is given by equation (\ref{G3}). 
The expressions involved in the calculations that follow this result become lengthy thus we shall not show them explicitly.
Just as in the previous cases, however, one can see that the procedure is straightforward.

\subsection{Painlev\'e V}
In \cite{gf01} the authors already studied the BT's of the fifth Painlev\'e equation 
by considering a solution $w=w(z;\alpha,\beta,\gamma,\delta)$ 
with parameters $\alpha$, $\beta$, $\gamma$, $\delta\neq0$, such that
\begin{equation}\nonumber
 F_1(w)=zw'-k_1cw^2+(k_1c-k_2a+k_3hz)w+k_2a\neq0,
\end{equation}
where  $c^2=2\alpha$, $a^2=-2\beta$, $h^2=-2\delta$. 
Then, they found that the transformation
\begin{equation}\nonumber
 T_{k_1,k_2,k_3}:\, w(z;\alpha,\beta,\gamma,\delta)\rightarrow w_1(z;\alpha_1,\beta_1,\gamma_1,\delta_1)=1-2k_3hzwF_1^{-1}
\end{equation}
defines another solution $w_1(z;\alpha_1,\beta_1,\gamma_1,\delta_1)$ of the PV equation with parameters
\begin{eqnarray}\nonumber
 \alpha_1=-\frac{1}{16\delta}\left[\gamma+k_3h(1-k_2a-k_1c)\right]^2,&& \beta_1=\frac{1}{16\delta}\left[\gamma-k_3h(1-k_2a-k_1c)\right]^2,\\\nonumber
 \gamma_1=k_3h(k_2a-k_1c),&& \quad\delta_1=\delta,
\end{eqnarray}
where $k_i^2=1$, $i\in{1,2,3}$. Thus $T_{k_1,k_2,k_3}$ is a BT for the PV equation.

Now, for a fixed value of $\epsilon$ in a first-order transformation and $\epsilon_1$ in a second-order transformation
such that $\epsilon_1=\epsilon$ we find that $T_{k_1,k_2,k_3}$ is a BT for the solutions found in Section \ref{PV}
in the following cases:
\begin{itemize}
 \item $T_{k_1,k_2,k_3}:w_{1b}\rightarrow w_{2a}$ whenever $k_1=-1$, $k_2=1$, $k_3=1$ if $\epsilon<-\frac{3}{2}$;
or $k_1=-1$, $k_2=-1$, $k_3=1$ if $-\frac{3}{2}<\epsilon<\frac{3}{2}$.

\item $T_{k_1,k_2,k_3}:w_{1b}\rightarrow w_{2e}$ whenever $k_1=-1$, $k_2=-1$, $k_3=1$ if $\epsilon<-\frac{3}{2}$;
or $k_1=-1$, $k_2=1$, $k_3=1$ if $-\frac{3}{2}<\epsilon<\frac{3}{2}$.

\item $T_{k_1,k_2,k_3}:w_{1c}\rightarrow w_{2a}$ whenever $k_1=1$, $k_2=-1$, $k_3=1$ and $\frac{1}{2}<\epsilon$.

\item $T_{k_1,k_2,k_3}:w_{1c}\rightarrow w_{2d}$ whenever $k_1=1$, $k_2=1$, $k_3=1$ if $\frac{1}{2}<\epsilon$;
$k_1=-1$, $k_2=1$, $k_3=1$ if either $-\frac{1}{2}<\epsilon<\frac{1}{2}$ or $\epsilon=\frac{1}{2}$;
or $k_1=-1$, $k_2=-1$, $k_3=1$ if $\epsilon<-\frac{1}{2}$.

\item $T_{k_1,k_2,k_3}:w_{1f}\rightarrow w_{2d}$ whenever $k_1=-1$, $k_2=-1$, $k_3=1$ and for all values of $\epsilon$.

\item $T_{k_1,k_2,k_3}:w_{1f}\rightarrow w_{2e}$ whenever $k_1=-1$, $k_2=1$, $k_3=1$ and for all values of $\epsilon$.

\item $T_{k_1,k_2,k_3}:w_{2d}\rightarrow w_{1c}$ whenever $k_1=1$, $k_2=1$, $k_3=-1$ if $\epsilon<-\frac{3}{2}$;
or $k_1=-1$, $k_2=-1$, $k_3=-1$ if $\frac{7}{2}<\epsilon$.

\item $T_{k_1,k_2,k_3}:w_{2d}\rightarrow w_{1f}$ whenever $k_1=-1$, $k_2=1$, $k_3=-1$ if $\epsilon<-\frac{3}{2}$;
$k_1=1$, $k_2=1$, $k_3=-1$ if $-\frac{3}{2}<\epsilon$; 
or $k_1=1$, $k_2=-1$, $k_3=-1$ if $\frac{7}{2}<\epsilon$

\item $T_{k_1,k_2,k_3}:w_{2e}\rightarrow w_{1b}$ whenever $k_1=1$, $k_2=1$, $k_3=-1$ if $\epsilon\leq-\frac{1}{2}$;
$k_1=-1$, $k_2=1$, $k_3=-1$ if $-\frac{1}{2}\leq\epsilon<\frac{5}{2}$;
or $k_1=-1$, $k_2=-1$, $k_3=-1$ if $\frac{5}{2}<\epsilon$.

\item $T_{k_1,k_2,k_3}:w_{2e}\rightarrow w_{1f}$ whenever $k_1=-1$, $k_2=1$, $k_3=-1$ if $\epsilon\leq-\frac{1}{2}$;
or $k_1=1$, $k_2=1$, $k_3=-1$ if $-\frac{1}{2}\leq\epsilon<\frac{5}{2}$.
\end{itemize}

\section{Conclusions}
This paper is a natural continuation of the work begun in \cite{fm14}, where non-singular SUSY transformations of 
first- and second-order were applied to the truncated oscillator and several solutions to the  Painlev\'e IV 
equation were obtained exploiting the connection between the algebraic structure of the supersymmetric partners of 
the truncated oscillator (second-order Polynomial Heisenberg Algebras) and the Painlev\'e IV equation. 
 
In this work we realize that one can also define fourth-order ladder operators for the SUSY partners of the truncated
oscillator in such a way that these systems are characterized as well by third-order Polynomial Heisenberg Algebras.
This fact was used to relate the extremal states of these Hamiltonians to several solutions of the Painlev\'e V equation,
resulting in a set of explicit expressions for such solutions in terms of a fixed parameter.

Finally, in order to gain a better insight of the relation among the solutions generated through said procedure,
we identified the B\"acklund transformations connecting the solutions we produced for the PIV and PV equations,
respectively. 
We have shown that Painlev\'e solutions generated by SUSY transformations of different orders are connected;
from the physical viewpoint this is unexpected since, in general, the spectra of the first-order SUSY partners 
of $H_0$ and those of second order are different.
We conclude that important new information for such SUSY generated solutions has been obtained in this paper.

\section{Acknowledgments}
The authors acknowledge the support of Conacyt (M\'exico).
VS Morales-Salgado also acknowledges the Conacyt fellowship 243374.

\end{document}